# Multifunctional graphene optical modulator and photodetector integrated on silicon waveguides

Nathan Youngblood[1], Yoska Anugrah[1], Rui Ma[1], Steven J. Koester[1], and Mo Li[1*]

[1]Department of Electrical and Computer Engineering, University of Minnesota, Minneapolis, MN 55455, USA

ABSTRACT

**For optical communication, information is converted between optical and electrical signal domains at a high rate. The devices to achieve such a conversion are various types of electro-optical modulators and photodetectors. These two types of optoelectronic devices, equally important, require different materials and consequently it has been challenging to realize both using a single material combination, especially in a way that can be integrated on the ubiquitous silicon platform. Graphene, with its gapless band structure, stands out as a unique optoelectronic material that allows both photodetection and optical modulation. Here, we demonstrate a single graphene-based device that simultaneously provides both efficient optical modulation and photodetection. The graphene device is integrated on a silicon waveguide and is tunable with a gate made from another layer of graphene to achieve near-infrared photodetection responsivity of 57 mA/W and modulation depth of 64%. This novel multifunctional device may lead to many unprecedented optoelectronic applications.**

[*] Corresponding author: moli@umn.edu

Optical communication systems are continuously being miniaturized to integrate a large number of previously discrete optoelectronic devices with silicon-based integrated circuits to achieve on-chip optical interconnects for high performance computation[1]. The integration of silicon photonics demands that multiple optoelectronic functionalities be achieved with fewer CMOS compatible materials in order to lower the cost without sacrificing performance. Chief among the building blocks of photonic systems are optical modulators and photodetectors[2-4]. These two types of devices operate based on very different mechanisms and consequently utilize different device geometries. They often have to be made of different materials that are difficult and costly to integrate with silicon photonics. Optical modulators are based on electro-optical or electro-absorptive effects in materials such as $LiNbO_3$, germanium and compound semiconductor heterostructures. In silicon photonics, the dispersion effect induced by carrier injection or depletion is the most common method used to achieve integrated optical modulation[5,6]. At the receiving end of optical links, photodetectors convert light back into electrical signals by absorbing photons and generating charges through photo-electric effects. Therefore, strong absorption and effective collection of photo-excited carriers are desired for efficient photodetection. Because of these distinctive requirements, to date, no device that can function as both a photodetector and a modulator, and whose role can be switched through external control, has been made with a single type of material. Such a simple yet multifunctional device, if implemented, not only can make integrated optical systems programmable and adaptable, but also can lead to novel applications such as optoelectronic oscillators and new schemes of optical computation and signal processing.

With its remarkable optical and electrical properties, graphene has been exploited as a multifunctional optoelectronic material to achieve a plethora of optoelectronic devices with high performance[7-9], including photodetectors[10-15], optical modulators[16,17], a polarizer[18] and saturable absorbers[19,20]. Most notably, graphene's optoelectronic properties, including absorption and dispersion, are highly tunable depending on its carrier concentration which can be controlled adaptively by electrostatic gating or lastingly by chemical doping[21,22]. For example, graphene optical modulators were demonstrated by utilizing the gate tuned interband absorption in graphene[16,17] and predicted to have very high speed and very low energy consumption[23,24]. At the same time, harnessing graphene's broadband and efficient interband transition, graphene

-2-

photodetectors have been demonstrated in both normal-incidence[10-12] and waveguide-integrated configurations[13-15]. Because of its two-dimensional structure, graphene is ideally suited for integration with planar photonic devices and the performance of the devices benefits significantly from the elongated optical interaction length in the coplanar configuration[13-15,25].

In this work, we fully utilize graphene's extraordinary and tunable optoelectronic properties to demonstrate the first optoelectronic device that acts as both a modulator and a photodetector, where the functionality of the device can be controlled with an integrated electrostatic gate also made of graphene. Fig. 1a illustrates the configuration of the device which consists of two layers of graphene, separated by a dielectric layer, and integrated on a planarized silicon photonic waveguide. The device is in the configuration of a simple field effect transistor (FET): the bottom layer (the channel) acts as an optical absorber and can collect photo-generated carriers while the top layer acts as a transparent gate electrode which can tune the electrical and optical properties of the bottom graphene layer. The graphene is grown by chemical vapor deposition (CVD) on copper foil and transferred onto the photonic waveguide substrate. The dielectric layer between the gate and the channel is 100 nm thick aluminum oxide ($Al_2O_3$) deposited by atomic layer deposition (ALD). The source and drain contacts on the channel are made of titanium/gold and palladium/gold, which have different work functions and dope graphene n-type and p-type[26-29], respectively. The differential metal-graphene contacts induce a lateral p-i-n junction, if the middle of the graphene channel is tuned to charge neutral as shown in Fig. 1b, with a built-in electrical field in the channel. This field facilitates the separation and drift of the photo-generated electrons and holes. This allows the device to generate a net photocurrent without the application of a bias voltage and with a higher efficiency than with a single-sided configuration[11,13]. Figs. 1c and 1d show the optical and scanning electron microscope images of a typical device. To most accurately measure the net optical absorption of graphene in order to determine the performance, the device is embedded in one arm of an unbalanced Mach-Zehnder interferometer. From the interference fringes, a comparison of the total optical absorption in the device arm can be made with respect to the reference arm without graphene and the absorption coefficient of graphene can be unambiguously determined[25]. A pair of grating couplers with efficiency of ~20% each are integrated with the interferometer to couple light in and out with optical fibers.

The FET configuration allows us to characterize the electrical properties of the graphene channel. Fig. 2a shows the resistance of the channel when the gate voltage is scanned. The results show that the charge neutrality point (CNP) is reached when a gate voltage of $V_g$=+33 V is applied, indicating that the graphene channel is heavily p-doped with a hole concentration of $p = 1.4 \times 10^{13}$ cm$^{-2}$ and a corresponding Fermi level of $E_F$=−0.45 eV. This level of doping is relatively high for graphene grown by CVD method and can be attributed to the trapped positive charges at the dielectric interface. Fitting the resistance vs. $V_g$ results, results in an extracted carrier mobility in the graphene of 1150 cm$^2$/V·s, which is relatively low and attributed to disorder introduced by Al$_2$O$_3$ deposition and charge trapping in the dielectric. Detailed modeling and analysis is included in the Supplementary Materials. Recent results of room temperature mobility up to 1.4×10$^5$ cm$^2$/V·s obtained from graphene on boron nitride[21] are very encouraging to improve the device performance demonstrated here. Fig. 2b shows the transmission spectrum of the Mach-Zehnder interferometer before the graphene layers were integrated on the waveguide. The interference fringes show an extinction ratio higher than 40dB ($ER=T_{max}/T_{min}$, $T_{max}$ and $T_{min}$ are the transmission at the peaks and valleys, respectively), confirming that there is negligible excess optical loss (less than 0.1dB) in the interferometer arms. During the fabrication of the device, the ER of the interferometer was measured after every step so that the optical loss contributed by each of the layers in the device can be accounted for. When the device was completed, the ER decreased to 1.6 when zero gate voltage was applied, corresponding to an added loss of 18 dB in the device arm. The major contribution to the optical loss is from the bottom layer of graphene that is directly above the waveguide, rather than from the top layer of graphene or the metal contact pads which are 1 μm away from the sides of the waveguide. When voltage was applied to the top graphene gate, the extinction ratio of the interference fringes was modulated, as shown in Fig. 2c. We observed that the ER increased (decreased) when negative (positive) gate voltage was applied, indicating reduced (augmented) absorption in the graphene. We measured the ER at every step of the applied gate voltage and calculated the linear absorption coefficient in the bottom graphene layer, as shown in Fig. 2d and e. The results show that the absorption coefficient can be modulated from a peak value of 0.2 dB/μm near CNP to 0.15 dB/μm, *i.e.* 25% of modulation, when $V_g$=−40V is applied. If the device is used as a modulator, the corresponding modulation depth of the optical intensity in the 90μm long waveguide is 64%. The change of absorption in graphene can be explained by the modulated Fermi level and the interband transition rate[21,22,30]. At $V_g=-40V$, the Fermi level in the channel graphene is expected to be $E_F=-0.69$ eV. Thus, for photon energy of 0.8 eV ($\lambda=1.55$ μm), interband absorption decreases as shown in Fig. 2e. Our theoretical analysis (in the supplementary materials) has taken into account the intraband absorption and the random potential fluctuations in the graphene and predicts that the absorption coefficient can be tuned to 5% of the peak value. However, in our devices, the measured absorption coefficient at room temperature remains at 75% of the value at CNP, leading to a high insertion loss of the modulator. One possible source of the higher-than expected absorption could be charge trapping in the $Al_2O_3$ dielectric that counteracts that action of gate voltage, particularly at very negative voltages. In addition, phonon-assisted mid-gap absorption in graphene has recently been suggested to contribute a significant amount (20-25%) of the total absorption in graphene and thus can be another source of the high residual absorption[31]. Both of these sources of absorption are expected to be significantly reduced by using boron nitride cladding layers for the graphene. Our results of gate modulated optical absorption in graphene also agree well with that of previous waveguide integrated graphene modulators[16,17]. We note that the interferometer configuration employed in this work unambiguously rules out the unvaried optical losses contributed from other parts of the device so that the tunable and residual parts of the absorption in the graphene can be determined and analyzed in more detail.

The strongly tunable optical absorption demonstrated above stems from the interband transition in graphene which at the same time generates photocarriers able to be collected so that both photodetection and optical modulation can be achieved with the same device. Such a multifunctional application has not been realized with any other optoelectronic materials. We measured the source-drain current when optical input was on (*i.e.* photocurrent) when the gate voltage was varied. As shown in Fig. 1b, the differential metal contacts induce a built-in internal field in the graphene channel, especially when the graphene is at the CNP, which facilitates the separation of photocarriers and generates a net photocurrent. Fig. 3a plots the measured photocurrent with zero source-drain bias and input optical power of 12 dBm (~1 mW in the waveguide) while the gate voltage was scanned. As expected, peak photocurrent was obtained very close to the CNP (the actual CNP drifts because of the slowly time-varying trapped charges at the graphene-dielectric interface). For comparison, the gate tuned absorption coefficient is also plotted in Fig. 3a. It appears that the peak of the photocurrent is much sharper than the absorption, indicating that in addition to modulating the absorption, the gate voltage also modulates the collecting efficiency of photocarriers by changing the field distribution in the channel. At the CNP, the built-in field can penetrate across the channel and the electron-electron scattering of the photocarriers is minimum. The sign of the photocurrent changes twice when the graphene is tuned from n-doped to heavily p-doped. This is consistent with the results from graphene photodetectors for normal incident light and can be explained by the contribution of photocurrent from both photovoltaic and photo-thermoelectric effects[32-35]. The photovoltaic current dominates when the doping in the graphene is low while the photo-thermoelectric current dominates at high doping regimes and has an opposite sign to the photovoltaic current. Detailed analysis and modeling of the photocurrent generation is included in the supplementary materials.

Knowing graphene's absorption coefficient $\alpha$ and thus the actual optical power absorbed in the graphene, the internal quantum efficiency $\eta$ of the photodetector can be determined by $\eta = I_p h\upsilon / eP_0(1-e^{-\alpha L})$, where $I_p$ is the photocurrent, $h\upsilon$ is the photon energy, $e$ is the electron charge, $P_0$ is the input optical power and $L$ is the length of the graphene detector. The result is plotted in Fig. 3b. A maximal total quantum efficiency of 0.25% and detector responsivity of 3.6 mA/W was obtained at the CNP. This photodetection responsivity is among the highest value achieved so far with single layer, CVD grown graphene. We note that although an internal efficiency of 6-16% was estimated in graphene under a high internal field which exists in the region very close to the metal contacts, the efficiency measured here is an averaged value across the 2.5 μm long graphene channel where the field is non-uniform and thus has a more practical meaning. To improve the quantum efficiency, it is essential to improve the efficiency of separating photocarriers which requires higher carrier mobility and stronger internal field. The former can be obtained with high quality graphene like what has been realized in 2D heterostructures of boron nitride and graphene[36] with reduced phonon-assisted scattering. Although zero-bias operation has zero dark current and is advantageous for low noise detection, application of a bias voltage between source and drain provides a stronger field in the channel which can significantly improve detection efficiency. Fig. 3c shows a 2D plot of the photocurrent measured with a different device when both gate and bias voltages were scanned. A six-fold sign change of the photocurrent is observed, consistent with previous results obtained in normal incidence devices[32]. At high doping and large bias, the photocurrent is dominated by the bolometric effect which has an opposite sign to the photovoltaic current which dominates at the low doping, low bias regime. In this device, the photovoltaic current dominated responsivity reaches 57 mA/W at a moderate bias voltage of 0.4V and near CNP with a gate voltage of 9 V, as shown in Fig. 3d and 3e.

We have demonstrated a novel multifunctional optoelectronic device based on graphene and integrated on a photonic waveguide that can be operated as both an optical modulator and a photodetector and can be tuned with a gate voltage. Further observation of Fig. 3a reveals that the optical absorption and the photocurrent are simultaneously modulated by the gate voltage. While the photocurrent should be proportional to the absorbed optical power and thus approximately the absorption coefficient, it is also sensitive to the field distribution in the graphene channel which is modulated by the gate. Thus the device can be operated in an unprecedented mode of simultaneous optical modulation and photodetection: the transmitted optical signal will be modulated by the gate voltage while the photocurrent measured at the same time is proportional to the absorbed optical power thereby its phase is opposite to the modulated optical signal. Thus this novel modulator-detector provides an in-situ feedback of the optical modulation which can, for example, be utilized to build an optoelectronic oscillator[37] with both electrical and optical outputs. In Fig. 4a, both the transmitted optical intensity and the photocurrent are modulated by a 100 kHz sinusoidal signal of 2V peak-to-peak amplitude applied to the gate. The optical transmission is clearly modulated out-of-phase with the simultaneously measured photocurrent. The speeds of graphene based modulators and photodetectors are expected to be very high with an intrinsic bandwidth estimated to approach 500 GHz[10] and 120 GHz[24], respectively. Fig. 4b and c show the measured frequency response of optical modulation and photodetection, respectively. The 3dB bandwidth of the device was measured to be 3 GHz for photodetection and 2.5 GHz for modulation, while our model predicts bandwidths of roughly 6 GHz and 9 GHz, respectively. In both cases, the main bandwidth limitation is attributed to the large parasitic capacitance arising from the contact pads and that bandwidths closer to the theoretical predictions should be possible using an optimized device geometry.

In conclusion, a multifunctional, waveguide integrated and gate tunable graphene optical modulator and photodetector has been demonstrated. Using CVD grown monolayer graphene, high modulation depth and detector responsivity have been achieved with a device of compact footprint. The performance is limited by the relatively low quality of the graphene that is used in the device and can be expected to improve dramatically by leveraging the latest research on heterostructures of graphene and other two-dimensional materials. Nevertheless, the demonstration that multiple optoelectronic functionalities can be obtained in a single device of one material is unprecedented and could open door to many future applications in integrated optoelectronics. In particular, the simplicity and the compatibility of the process to integrate graphene are very promising for large-scale integrated silicon photonics.

**Method:**

The dual-layer, graphene modulator/detector was fabricated using silicon-on-insulator (SOI) wafers (SOITEC Corp.) with 110 nm top silicon layer and 3 μm buried oxide layer. The underlying photonics layer was patterned using electron beam lithography and plasma etching. Electron beam evaporation was then used to deposit 145 nm silicon dioxide on the sample using the remaining negative ebeam resist as a self-aligned mask. The planarized substrate was annealed using rapid thermal annealing at 1100 ºC for 1 minute to improve the quality of the evaporated oxide. CVD graphene (purchased from Graphene Square Inc.) was then transferred using a typical PMMA assisted transfer method. Ebeam lithography followed by an oxygen plasma etch was used to pattern graphene mesas. Source (drain) contacts were subsequently defined with ebeam lithography and evaporated 20 nm Pd (Ti)/ 80 nm Au. 100 nm aluminum oxide was deposited as a gate oxide using atomic layer deposition following a 1.5 nm evaporated aluminum seed layer. Contact holes were etched through the gate oxide using buffered oxide etch. The top graphene gate was transferred and defined using the same techniques mentioned above. Finally, top gate contacts of 5 nm Ti / 100 nm Au were patterned using ebeam lithography and lift-off.

Measurement of light modulation was achieved by recording the device spectrum (from 1520 nm to 1620 nm) at various gate voltages and performing a fit to the transmission peaks and valleys of the MZ structures. Waveguide loss was also measured in this way after the fabrication of each device layer. With this method, absorption due to the contacts and top gate was subtracted from the variable absorption in the active graphene layer. Photocurrent measurements were obtained by connecting the source / drain contacts to a current preamplifier and lock-in amplifier. A 10 kHz signal from the lock-in was used to modulate the laser source illuminating the device. Photocurrent measured by the lock-in was recorded while scanning the gate voltage from 40V to -40V. The internal quantum efficiency of the device was then extracted using the measured gate-dependent photocurrent and graphene absorption. Photocurrent maps at different bias and gate voltages were measured at a fixed wavelength for various laser powers. Two source-meters (Keithley 2400) were used in this measurement, one to control the gate voltage and the other to bias the device while measuring the source-drain current. The devices were held at a fixed bias while the gate voltage was swept twice – once with no optical power and once with the laser source on. The difference between the two scans yielded gate and bias dependent photocurrent. Power and bias dependent photocurrent was extracted from these two dimensional scans and responsivity was calculated after accounting for loss due to the input grating coupler and power splitting arm of the MZ structure. Simultaneous time dependent measurements of the photocurrent and light modulation were performed by sending CW light to the device while modulating the gate voltage with a 100 kHz / 2 $V_{pp}$ sine wave at -1V DC offset. The photocurrent and transmission traces were averaged and captured on an oscilloscope. Frequency dependent photocurrent was measured by modulating the laser source with an optical modulator (Lucent 2623NA) driven by a vector network analyzer (Anritsu 37369D). The modulated source was amplified with a Pritel erbium doped fiber amplifier (EDFA) and sent to the device under testing. The source / drain electrodes were connected to a low noise pre-amplifier and fed back into the network analyzer. For frequency dependent modulation, CW illumination was coupled to the device and the transmission was monitored with a 12 GHz photoreceiver (Newport 1554-A). The network analyzer was used to drive the gate voltage and measure the frequency response from the photoreceiver.

**Acknowledgements**

M.L. acknowledges the funding support provided by the Young Investigator Program (YIP) of AFOSR (Award No. FA9550-12-1-0338). S.J.K and Y.A. would like to acknowledge partial support of NSF and NRI under NSF Grant No. ECCS-1124831. Parts of this work was carried out in the University of Minnesota Nanofabrication Center which receives partial support from NSF through NNIN program, and the Characterization Facility which is a member of the NSF-funded Materials Research Facilities Network via the MRSEC program.

**Author Contributions**

N.Y. fabricated the devices, performed the measurements and analyzed the data. Y.A. conducted and R.M. assisted the transfer of the graphene. M.L. and S.J.K. conceived and supervised the research. M.L., S.J.K. and N.Y. co-wrote the manuscript.

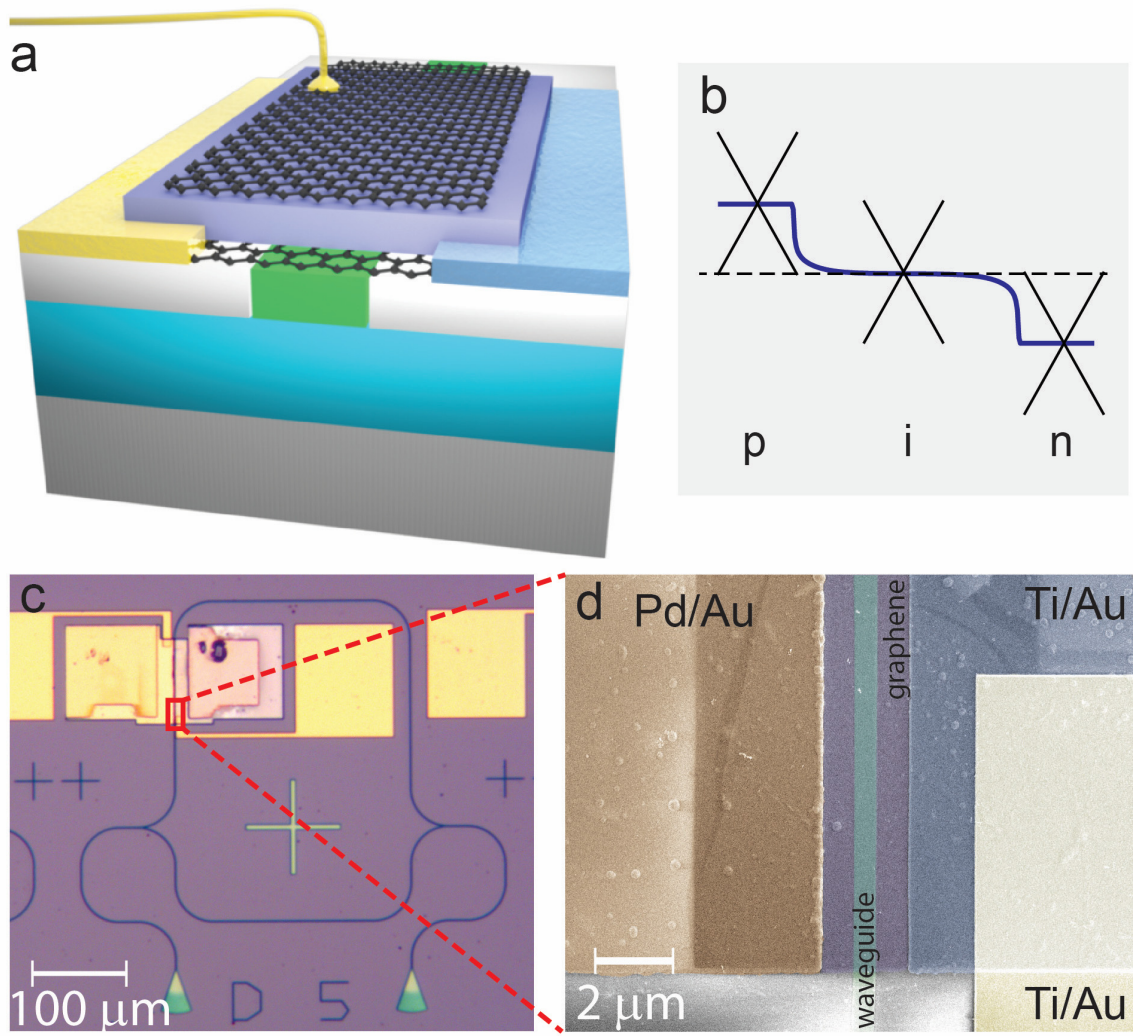

**Figure 1 Waveguide integrated graphene modulator and photodetector. a)** Schematic illustration of the dual layer graphene modulator/detector integrated on a planarized waveguide. **b)** Illustration of the profile across the graphene channel highlighting the p-i-n junction and the build-in electric field. **c)** Optical microscope image of the device. A Mach-Zehnder interferometer made of silicon waveguides is employed to accurately determine the optical absorption in graphene. Grating couplers (green triangles at the bottom) are used to couple light in and out of the device. **d)** Scanning electron microscope image of the device. Two asymmetric source and drain contacts made of titanium/gold (light blue) and palladium/gold (dark yellow) dope the graphene (light purple) to be n- and p-type, respectively. Contact (light yellow) to the top graphene gate is made of titanium/gold. The waveguide is in light green.

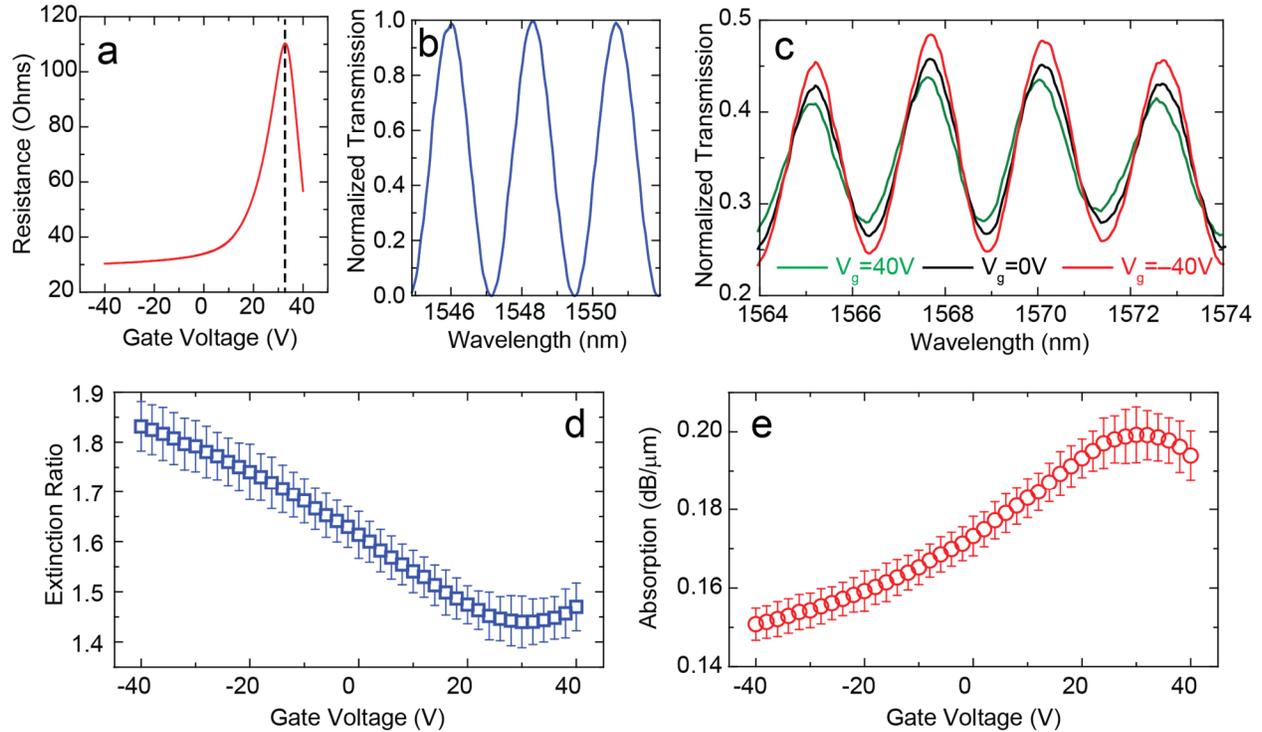

**Figure 2 Gate tuned optical absorption and optical modulation. a)** Source-drain resistance of the graphene channel versus the voltage applied to the top graphene gate. The charge neutral point (CNP) is reach at +33 V, indicating the graphene as-transferred is highly p-type. **b)** Interference fringes measured from the silicon waveguide Mach-Zehnder interferometer just before the graphene layers were transferred. Very high extinction ratio (>40dB) indicates very low optical loss (<0.1dB) in the waveguide. **c)** Interference fringes measured after the graphene device was completed and different gate voltages were applied. The extinction ratio decreased (increased) when +(−)40V was applied. **d)** Measured interference extinction ratio at various gate voltages. **e)** The absorption coefficient in the graphene channel calculated from the extinction ratio at various gate voltages and excluding absorption in other parts (top graphene gate and metal contacts) of the devices. The absorption coefficient can be modulated from 0.2 dB/μm at $V_g$=+33V to 0.15 dB/μm at $V_g$=−40V. The corresponding modulation depth of the transmission in the waveguide is 64%.

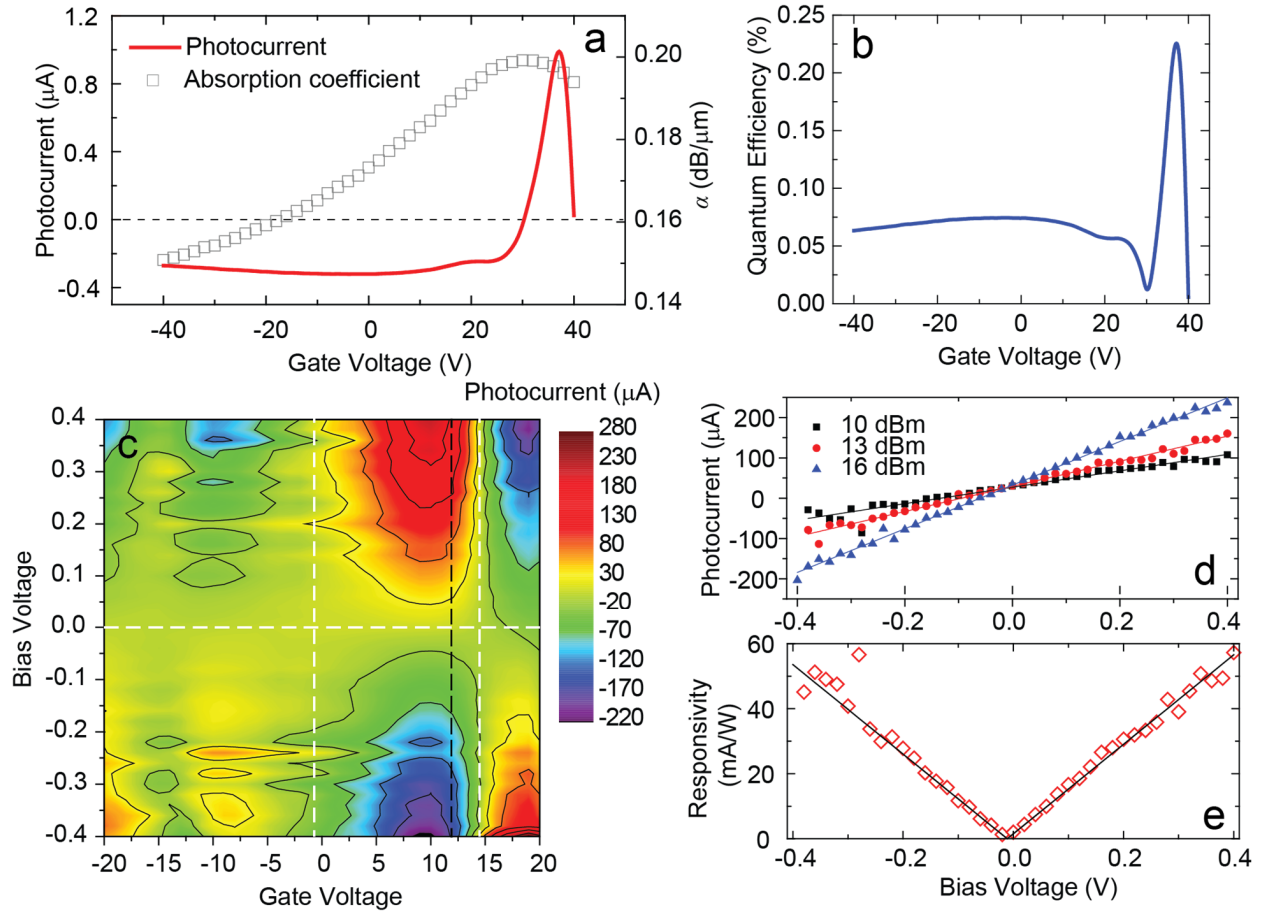

**Figure 3 Gate tuned photodetection with graphene. a)** Measured zero-bias photocurrent and graphene absorption coefficient at various gate voltages. The photocurrent peaks at the charge neutral point and changes sign twice when the gate voltage is scanned from negative to positive, indicating both photovoltaic and photo-thermoelectric contributions. **b)** Calculated quantum efficiency of the graphene photodetector. Maximum efficiency of 0.25% was reached at CNP. **c)** Two-dimensional plot of photocurrent measured from a different device when both gate and bias voltages are scanned. A six-fold change of the photocurrent direction can be observed. This can be explained by the dominance of bolometric effect in high doping regime and photovoltaic effect in the low doping regime, respectively. Black dashed line marks the CNP and white dashed lines are guides to the eyes demarcating the regions of positive and negative photocurrent. **d)** Photocurrent versus bias voltage at various input optical power and **e)** detector responsivity versus bias voltage measured with the gate voltage at +9 V near CNP. A maximal responsivity of 57 mA/W was measured at a moderate bias voltage of 0.4V.

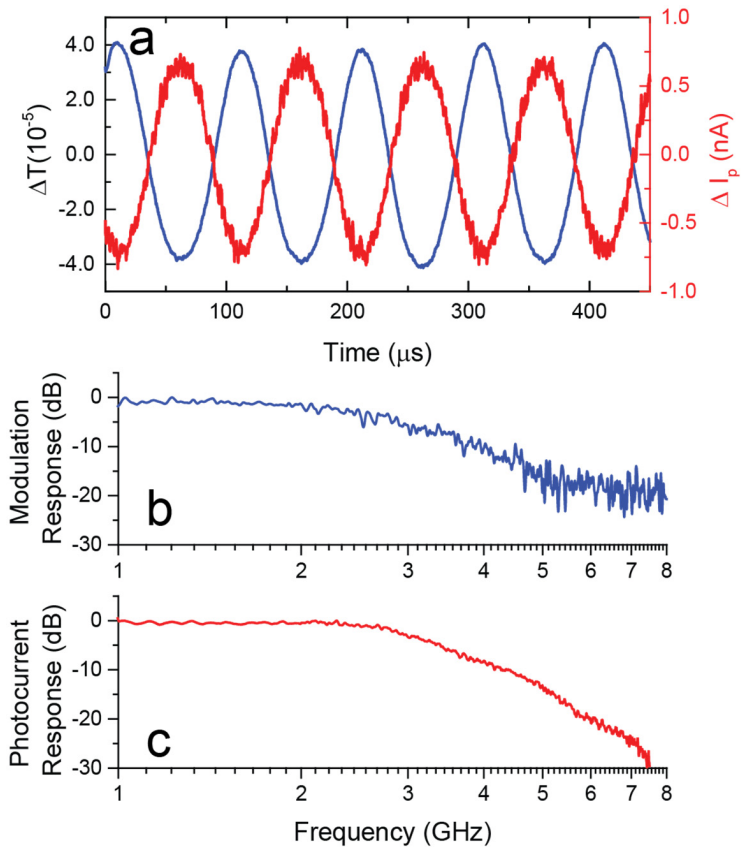

**Figure 4 Simultaneous optical modulation and photodetection. a)** Transmitted optical signal (blue trace) and photocurrent (red) measured simultaneously when a 100 kHz sinusoidal voltage of 2V peak-to-peak was applied to the gate. It is clear the photocurrent modulation, which should be proportional to optical absorption in graphene, is out-of-phase with the optical transmission. **b)** and **c)** Frequency response of optical modulation (b) and photocurrent (c). The 3dB bandwidth of the device is determined to be 3GHz for photodetection and 2.5 GHz for modulation. Both are limited by the RC time constants of the contact pads and gate capacitance in the device.